\documentclass[11pt]{article}
\usepackage{graphicx}
\usepackage{amsmath}
\usepackage{amssymb}
\usepackage{caption2}
\begin{document}
\bibliographystyle{prsty}
\begin{center}
{\large {\bf \sc{  Analysis of the 1S and 2S states of the  $\Lambda_Q$ and $\Xi_Q$  with the QCD sum rules }}} \\[2mm]
Zhi-Gang Wang \footnote{E-mail:zgwang@aliyun.com.}, Hui-Juan Wang      \\
 Department of Physics, North China Electric Power University,
Baoding 071003, P. R. China

\end{center}

\begin{abstract}
In this article, we  study  the ground states and the first radial excited states of the  flavor antitriplet heavy baryon states $\Lambda_Q$ and $\Xi_Q$ with the spin-parity $J^P={1\over 2}^{+}$ by carrying  out the operator product expansion up to the vacuum condensates of dimension $10$ in a consistent  way.  We observe that the higher  dimensional vacuum condensates play an important role, and obtain very stable QCD sum rules with variations of the Borel parameters for the heavy baryon states  for the first time.  The predicted masses $6.08\pm0.09\,\rm{GeV}$, $2.78\pm0.08\,\rm{GeV}$ and $2.96\pm0.09\,\rm{GeV}$ for the first radial excited states $\Lambda_b(\rm 2S)$, $\Lambda_c(\rm 2S)$ and $\Xi_c(\rm 2S)$ respectively are in excellent agreement with the experimental data  and support  assigning the $\Lambda_b(6072)$, $\Lambda_c(2765)$ and $\Xi_c(2980/2970)$ to be the first radial excited states of the $\Lambda_b$, $\Lambda_c$ and $\Xi_c$, respectively, the predicted mass $6.24\pm0.07\,\rm{GeV}$ for the $\Xi_b(\rm 2S)$ can be confronted to the experimental data in the future.
\end{abstract}



\section{Introduction}

Recently, the CMS collaboration observed a broad excess of events in the region of $6040-6100\,\rm{MeV}$ in the $\Lambda_b^0\pi^+\pi^-
$ invariant  mass spectrum based on a data sample  corresponding to an integrated luminosity of up to $140\, \rm{fb}^{-1}$ \cite{CMS-6072}. If it is fitted  with a single Breit-Wigner function, the obtained mass and width are $M= 6073\pm 5\,\rm{MeV}$ and $\Gamma = 55 \pm 11\,\rm{MeV}$, respectively.
Subsequently, the LHCb collaboration observed  a new excited baryon  state  in the $\Lambda_b^0\pi^+\pi^-
$  invariant  mass spectrum with high significance using a data sample corresponding to an integrated luminosity of $\rm 9\,fb^{-1}$. The measured  mass and natural width  are $M= 6072.3\pm 2.9\pm 0.6\pm 0.2\,\rm{MeV}$ and $\Gamma = 72\pm 11\pm2\,\rm{MeV}$, respectively, which are consistent with the first radial excitation of the $\Lambda_b^0$   baryon, the $\Lambda_b^0(\rm 2S)$  resonance \cite{LHCb-6072}. The $\Lambda_b(6072)$ can be assigned to be the $\Lambda_b^0(\rm 2S)$ state \cite{Lambdab-2S}, or  assigned to be the lowest $\rho$-mode excitation in $\Lambda_b$  family \cite{LuQF-6072}.

In 2001, at the charm sector, the CLEO collaboration observed the $\Lambda_c^+(2765)$ or $\Sigma_c^+(2765)$ in the $\Lambda_c^+ \pi^- \pi^+$ invariant mass spectrum using a $13.7 \rm fb^{-1}$  data sample recorded by the CLEO detector at CESR \cite{CLEO-2765}. The Belle  collaboration
determined   the isospin of the $\Lambda_c^+(2765)$ or $\Sigma_c^+(2765)$ to be zero  using a $980\rm fb^{-1}$  data sample in the $e^+ e^- $
 annihilation around   $\sqrt{s} =10.6\,\rm{GeV}$, and established it to be a $\Lambda_c$ resonance \cite{Belle-2765}. The $\Lambda_c(2765)$ can be assigned to be the $\Lambda_c(\rm 2S)$ state \cite{ChenB-2765-2980,Ebert-2765-2980}, however,  there are  several other possible  assignments \cite{ChengHY-review-2015}.

In 2006, the Belle collaboration reported the first observation of two charmed strange baryon states that decay into the final state $\Lambda_c^+ K^-\pi^+
$, the broader one has a mass of $2978.5\pm 2.1\pm 2.0\, \rm{ MeV}$   and a width of $43.5\pm 7.5\pm 7.0 \,\rm{ MeV}$ \cite{Belle-2980}.  Subsequently, the BaBar collaboration confirmed the $\Xi_c(2980)$ or $\Xi_c(2970)$ \cite{BaBar-2980}.  The $\Xi_c(2980/2970)$ can be assigned to be the $\Xi_c(\rm 2S)$ state \cite{ChenB-2765-2980,Ebert-2765-2980}, however, there are  several other possible  assignments \cite{ChengHY-review-2015}.

The mass spectrum of the single heavy baryon states has been studied intensively in various theoretical models  \cite{Lambdab-2S,LuQF-6072,ChenB-2765-2980,Ebert-2765-2980,ChengHY-review-2015,Roberts-2008,Bagan92,Chung82,M-Nielsen07,ZhangJR-2008,WangZG-EPJC-2010,WangZG-PLB,Omega-c}.
If the $\Lambda_b(6072)$, $\Lambda_c(2765)$ and $\Xi_c(2980)$ are the first radial excited states of the $\Lambda_b$, $\Lambda_c$ and $\Xi_c$, respectively,
the mass gaps between the ground states and the first radial excited states are less than $0.5\,\rm{GeV}$, which are significantly lower than the amount that is expected by the  3-dimensional harmonic oscillator model. In the QCD sum rules for the single heavy baryon states, if we carry out the operator product expansion up to the vacuum condensates of dimension 6, we have to choose the continuum threshold parameters as $\sqrt{s_0}=M_{gr}+0.6\sim0.8\,\rm{GeV}$ or $0.7\sim0.9\,\rm{GeV}$ to reproduce the experimental data \cite{M-Nielsen07,ZhangJR-2008,WangZG-EPJC-2010,WangZG-PLB}, where the subscript $gr$ stands for the ground states.
The energy gaps $0.6\sim0.8\,\rm{GeV}$ and $0.7\sim0.9\,\rm{GeV}$ are much larger than the physical energy gap $0.5\,\rm{GeV}$, the contributions of the first radial excited states are included in.  The heavy baryon states, which have one heavy quark  and two light quarks, play an important  role in understanding the dynamics of light quarks in the presence of one heavy quark, also in  understanding of the confinement mechanism  and the heavy quark symmetry.

At the hadron side of the correlation functions in the QCD sum rules for the heavy baryon states, there are one heavy quark propagator and two light quark propagators. If the heavy quark line emits a gluon, each light quark line contributes a quark-antiquark pair, we obtain quark-gluon operators of dimension 10. In previous works, the operator product expansion was  carried out up to the vacuum condensates of dimension 6 \cite{Bagan92,Chung82,M-Nielsen07,ZhangJR-2008,WangZG-EPJC-2010,WangZG-PLB}. In Ref.\cite{WangZG-EPJC-2010}, we study the masses and pole residues of the ${1\over 2}^\pm$ flavor antitriplet heavy baryon  states
($\Lambda_c^+$, $\Xi_c^+,\Xi_c^0)$ and ($\Lambda_b^0$, $\Xi_b^0,\Xi_b^-)$ by subtracting the contributions from the
corresponding ${1\over 2}^\mp$ heavy baryon  states with the QCD sum rules. Now we revisit our previous work by calculating the vacuum condensates up to dimension 10, and extend our previous work to study  the first radial excited states $\Lambda_Q(\rm 2S)$ and $\Xi_Q(\rm 2S)$, and make possible assignments of the  $\Lambda_b(6072)$, $\Lambda_c(2765)$ and $\Xi_c(2980)$.

The article is arranged as follows:  we derive the QCD sum rules for the masses and the pole residues of  the heavy baryon states
$\Lambda_Q(\rm 1S,2S)$ and $\Xi_Q(\rm 1S, 2S)$ in Sect.2; in Sect.3, we present the  numerical results and discussions; and Sect.4 is reserved for our
conclusions.

\section{QCD sum rules for  the $\Lambda_Q(\rm 1S,2S)$ and $\Xi_Q(\rm 1S, 2S)$ }
We interpolate the spin-parity $J^P={1\over 2}^+$ flavor  antitriplet heavy baryon states
$\Lambda_Q$, $\Lambda_Q(\rm 2S)$, $\Xi_Q$  and $\Xi_Q(\rm 2S)$ with the $\Lambda$-type currents $J_\Lambda(x)$ and $J_\Xi(x)$,
 respectively,
\begin{eqnarray}
J_\Lambda(x)&=& \varepsilon^{ijk}  u^T_i(x)C\gamma_5 d_j(x)   Q_k(x)  \, ,  \nonumber \\
J_\Xi(x)&=& \varepsilon^{ijk}  q^T_i(x)C\gamma_5 s_j(x)   Q_k(x)  \, ,
\end{eqnarray}
where  $Q=c$, $b$, $q=u$, $d$, the $i$, $j$ and $k$ are color indexes, and the $C$ is the charge conjunction matrix.

The attractive interaction induced by one-gluon exchange  favors  forming the diquark states or quark-quark-correlations in the  color antitriplet $\overline{3}_{ c}$ \cite{One-gluon}.
The color antitriplet  diquark operators $\varepsilon^{ijk} q^{T}_j C\Gamma q^{\prime}_k$
  have  five  structures  in the  Dirac spinor space, where $C\Gamma=C\gamma_5$, $C$, $C\gamma_\mu \gamma_5$,  $C\gamma_\mu $ and $C\sigma_{\mu\nu}$ for the scalar, pseudoscalar, vector, axialvector  and  tensor diquarks, respectively, and couple potentially to the corresponding scalar, pseudoscalar, vector, axialvector  and  tensor diquark states, respectively.
The calculations via the QCD sum rules  indicate that  the favored quark-quark configurations are the scalar and axialvector diquark states, while the most favored  quark-quark configurations are the scalar    diquark states \cite{WangLDiquark}.
We usually resort to the light-diquark-heavy-quark model to study the heavy baryon states. In the diquark-quark models, the angular momentum between the two light quarks is denoted as $L_\rho$, while the angular momentum between the  light diquark and the heavy quark is denoted as $L_\lambda$. If the two light quarks in the  diquark are in relative S-wave or $L_\rho=0$, then the heavy baryon states with the spin-parity  $J^P=0^+$ and $1^+$ diquark constituents are called $\Lambda$-type and $\Sigma$-type baryons, respectively \cite{Korner-PPNP}. In this article, we study the ground states and the first radial excited states of the $\Lambda$-type heavy baryons with the $\Lambda$-type interpolating currents.

We can interpolate the corresponding spin-parity $J^P={1\over 2}^-$ flavor antitriplet heavy baryon states
 with the $\Lambda$-type currents $i \gamma_{5}J_\Lambda(x)$ and $i \gamma_{5}J_\Xi(x)$ without introducing the relative P-wave explicitly,  because
multiplying $i \gamma_{5}$ to the currents $J_\Lambda(x)$ and $J_\Xi(x)$ changes their parity  \cite{Oka96}. Now let us write down the correlation functions,
\begin{eqnarray}
\Pi(p)&=&i\int d^4x e^{ip \cdot x} \langle0|T\Big\{J(x)\bar{J}(0)\Big\}|0\rangle \, ,
\end{eqnarray}
where $J(x)=J_\Lambda(x)$ and $J_\Xi(x)$.

We  insert  a
complete set  of intermediate baryon states with the same quantum numbers as the current operators $J_\Lambda(x)$, $i \gamma_{5}J_\Lambda(x)$, $J_\Xi(x)$  and
$i \gamma_{5}J_\Xi(x)$ into the
correlation functions $\Pi(p)$  to obtain the hadronic representation \cite{SVZ79,Reinders85}. After isolating the pole terms of
the ground states and the first radial excited  states, we obtain the following results,
\begin{eqnarray}
  \Pi(p) & = &   \lambda_+^2 {\!\not\!{p} +    M_{+} \over M^{2}_+ -p^{2} } +  \lambda_{\rm 2S,+}^2 {\!\not\!{p} +    M_{\rm 2S,+} \over M^{2}_{\rm 2S,+} -p^{2} } +\lambda_{-}^2  {\!\not\!{p} - M_{-} \over M_{-}^{2}-p^{2}  }+\lambda_{\rm 2S,-}^2  {\!\not\!{p} - M_{\rm 2S,-} \over M_{\rm 2S,-}^{2}-p^{2}  } +\cdots \, ,
    \end{eqnarray}
where the $M_{\pm}$ and $M_{\rm 2S,\pm}$ are the masses of the ground states and the first radial excited states with the parity
$\pm$ respectively, and the $\lambda_{\pm}$ and $\lambda_{\rm 2S,\pm}$ are the  corresponding
pole residues defined by $\langle 0|J(0)|B_{\pm/\rm 2S,\pm}(p)\rangle=\lambda_{\pm/\rm 2S,\pm}$,  $B=\Lambda_Q$ and $\Xi_Q$.

We rewrite the correlation functions as
\begin{eqnarray}
    \Pi(p) = \!\not\!{p}\, \Pi_{1}(p^2) + \Pi_{0}(p^2)\, ,
\end{eqnarray}
according to the Lorentz covariance, and obtain the hadronic spectral densities through dispersion relation,
\begin{eqnarray}
\rho_{H,1}(s)&=&{\rm lim}_{\epsilon \to 0}\frac{{\rm{Im}}\Pi_1(s+i\epsilon) }{\pi}\, , \nonumber\\
&=&\lambda_+^2 \delta\left(s- M^{2}_+\right) +  \lambda_{\rm 2S,+}^2\delta\left(s- M^{2}_{\rm 2S,+} \right) +\lambda_{-}^2 \delta\left(s- M^{2}_{-}\right) +  \lambda_{\rm 2S,-}^2\delta\left(s- M^{2}_{\rm 2S,-} \right) \nonumber\\
&&+ \cdots \, ,\\
\rho_{H,0}(s)&=&{\rm lim}_{\epsilon \to 0}\frac{{\rm{Im}}\Pi_0(s+i\epsilon) }{\pi}\, , \nonumber\\
&=&M_+\lambda_+^2 \delta\left(s- M^{2}_+\right) +  M_{\rm 2S,+}\lambda_{\rm 2S,+}^2\delta\left(s- M^{2}_{\rm 2S,+} \right) -M_{-}\lambda_{-}^2 \delta\left(s- M^{2}_{-}\right) \nonumber\\
&&-  M_{\rm 2S,-} \lambda_{\rm 2S,-}^2\delta\left(s- M^{2}_{\rm 2S,-} \right)+ \cdots \, ,
\end{eqnarray}
where we add the subscript $H$ to denote the hadron side of the correlation functions.

Now we carry out the operator product expansion up to the vacuum condensates of dimension 10 in a consistent way, and take into account the vacuum condensates which are quark-gluon operators of the order $\mathcal{O}(\alpha_s^k)$ with $k\leq1$. Again, we obtain the corresponding QCD spectral densities through dispersion relation,
  \begin{eqnarray}
\rho_{QCD,1}(s)&=&{\rm lim}_{\epsilon \to 0}\frac{{\rm{Im}}\Pi_1(s+i\epsilon) }{\pi}\, , \nonumber\\
\rho_{QCD,0}(s)&=&{\rm lim}_{\epsilon \to 0}\frac{{\rm{Im}}\Pi_0(s+i\epsilon) }{\pi}\, ,
\end{eqnarray}
where we add the subscripts $QCD$ to denote the QCD side of the correlation functions.

Then we choose the continuum thresholds $s_0$ and $s_0^\prime$ to include the ground states and the ground states plus the first radial excited states, respectively, and introduce the weight function $\exp\left(-\frac{s}{T^2}\right)$ to suppress the contributions of the higher resonances and continuum states. We take the combination,
\begin{eqnarray}
\int_{m_Q^2}^{s_0/s_0^\prime}ds \Big[\sqrt{s}\rho_{H,1}(s)+\rho_{H,0}(s)\Big]\exp\left(-\frac{s}{T^2}\right)\, ,
\end{eqnarray}
to exclude the contaminations  from the heavy baryon states with the negative parity,  and match the hadron side with the QCD side of the correlation functions. The  combinations,
\begin{eqnarray}
\int_{m_Q^2}^{\infty}ds \Big[\sqrt{s}\rho_{H,1}(s)\pm\rho_{H,0}(s)\Big]\exp\left(-\frac{s}{T^2}\right)\, ,
\end{eqnarray}
pick up the  heavy baryon states with the positive parity and negative parity, respectively.

Finally, we obtain  two QCD sum rules,
\begin{eqnarray}\label{QCDSR-I}
  2M_{+}\lambda_{+}^2\exp\left(-\frac{M_{+}^2}{T^2}\right)&=&\int_{m_Q^2}^{s_0}ds \Big[\sqrt{s}\rho_{H,1}(s)+\rho_{H,0}(s)\Big]\exp\left(-\frac{s}{T^2}\right) \, ,\nonumber\\
  &=&\int_{m_Q^2}^{s_0}ds \Big[\sqrt{s}\rho_{QCD,1}(s)+\rho_{QCD,0}(s)\Big]\exp\left(-\frac{s}{T^2}\right) \, ,
\end{eqnarray}

\begin{eqnarray}\label{QCDSR-II}
&& 2M_{+} \lambda_{+}^2\exp\left(-\frac{M_{+}^2}{T^2}\right)+2M_{\rm 2S,+}\lambda_{\rm 2S,+}^2\exp\left(-\frac{M_{\rm 2S,+}^2}{T^2}\right)\nonumber\\
&=&\int_{m_Q^2}^{s_0^\prime}ds \Big[\sqrt{s}\rho_{H,1}(s)+\rho_{H,0}(s)\Big]\exp\left(-\frac{s}{T^2}\right) \, ,\nonumber\\
  &=&\int_{m_Q^2}^{s_0^\prime}ds \Big[\sqrt{s}\rho_{QCD,1}(s)+\rho_{QCD,0}(s)\Big]\exp\left(-\frac{s}{T^2}\right) \, ,
\end{eqnarray}
where $\rho_{QCD,1}(s)=\rho_{\Lambda,1}(s)$, $\rho_{\Xi,1}(s)$, $\rho_{QCD,0}(s)=m_Q\,\rho_{\Lambda,0}(s)$, $m_Q\,\rho_{\Xi,0}(s)$,
\begin{eqnarray}
\rho_{\Lambda,1}(s)&=&\rho_{\Xi,1}(s)\mid_{m_s\to 0,\, \langle\bar{s}s\rangle\to\langle\bar{q}q\rangle, \,\langle\bar{s}g_s\sigma Gs\rangle\to\langle\bar{q}g_s\sigma Gq\rangle }\, , \nonumber \\
\rho_{\Lambda,0}(s)&=&\rho_{\Xi,0}(s)\mid_{m_s\to 0, \, \langle\bar{s}s\rangle\to\langle\bar{q}q\rangle, \,\langle\bar{s}g_s\sigma Gs\rangle\to\langle\bar{q}g_s\sigma Gq\rangle }\, ,
\end{eqnarray}

\begin{eqnarray}
\rho_{\Xi,1}(s)&=&\frac{3}{128\pi^4}\int_{x_i}^1dx \,x(1-x)^2(s-\widetilde{m}_Q^2)^2+\frac{ m_s [\langle \bar{s}s\rangle-2\langle \bar{q}q\rangle]}{32\pi^2}\left(1-x_i^2 \right)  \nonumber\\
 &&-\frac{m_s[\langle \bar{s}g_s\sigma Gs\rangle-3\langle \bar{q}g_s\sigma Gq\rangle]}{96\pi^2}\delta (s-m_Q^2) +\frac{\langle \bar{s}s\rangle\langle \bar{q}q\rangle}{6}\delta (s-m_Q^2) \nonumber\\
 &&-\frac{[\langle \bar{s}s\rangle\langle \bar{q}g_s\sigma Gq\rangle+\langle \bar{s}g_s\sigma Gs\rangle\langle \bar{q}q\rangle]}{24 T^2}\left(1+\frac{s}{T^2} \right)\delta (s-m_Q^2)  \nonumber\\
  &&+\frac{m_Q^4\langle \bar{q}g_s\sigma Gq\rangle\langle \bar{s}g_s\sigma Gs\rangle}{96 T^8}\delta (s-m_Q^2) +\frac{1}{256\pi^2}\langle\frac{\alpha_{s}GG}{\pi}\rangle\left(1-x_i^2 \right) \nonumber\\
&&-\frac{m_{Q}^{2}}{384\pi^2} \langle\frac{\alpha_{s}GG}{\pi}\rangle  \int_{x_{i}}^{1}dx\, \frac{(1-x)^2}{x^2}-\frac{m_sm_Q^2[\langle \bar{s}s\rangle-2\langle \bar{q}q\rangle]}{288T^4}\langle\frac{\alpha_{s}GG}{\pi}\rangle \frac{1-x_i}{x_i}\nonumber\\
&&-\frac{m_Q^2\langle \bar{s} s\rangle\langle \bar{q}q\rangle\pi^2}{108T^6}\langle\frac{\alpha_{s}GG}{\pi}\rangle\delta (s-m_Q^2)\, ,
\end{eqnarray}

\begin{eqnarray}
\rho_{\Xi,0}(s)&=&\frac{3}{128\pi^4}\int_{x_i}^1dx \,(1-x)^2(s-\widetilde{m}_Q^2)^2+\frac{ m_s [\langle \bar{s}s\rangle-2\langle \bar{q}q\rangle]}{16\pi^2}\left(1-x_i \right) \nonumber\\
 &&-\frac{m_s[\langle \bar{s}g_s\sigma Gs\rangle-3\langle \bar{q}g_s\sigma Gq\rangle]}{96\pi^2}\delta (s-m_Q^2) +\frac{\langle \bar{s}s\rangle\langle \bar{q}q\rangle}{6}\delta (s-m_Q^2) \nonumber\\
 &&-\frac{m_Q^2[\langle \bar{s}s\rangle\langle \bar{q}g_s\sigma Gq\rangle+\langle \bar{s}g_s\sigma Gs\rangle\langle \bar{q}q\rangle]}{24 T^2}\delta (s-m_Q^2)  \nonumber\\
  &&+\frac{m_Q^2\langle \bar{q}g_s\sigma Gq\rangle\langle \bar{s}g_s\sigma Gs\rangle}{48 T^6}\left(-1+\frac{s}{2T^2} \right)\delta (s-m_Q^2)  \nonumber\\
&&-\frac{m_{Q}^{2}}{384\pi^2} \langle\frac{\alpha_{s}GG}{\pi}\rangle  \int_{x_{i}}^{1}dx\, \frac{(1-x)^2}{x^3}+\frac{1}{128\pi^2}\langle\frac{\alpha_{s}GG}{\pi}\rangle\int_{x_{i}}^{1}dx\,\frac{(1-x)^2}{x^2}\nonumber\\
&&+\frac{1}{128\pi^2}\langle\frac{\alpha_{s}GG}{\pi}\rangle\left(1-x_i \right)-\frac{m_sm_Q^2[\langle \bar{s}s\rangle-2\langle \bar{q}q\rangle]}{576T^4}\langle\frac{\alpha_{s}GG}{\pi}\rangle \,\frac{1-x_i^2}{x_i^2}\nonumber\\
&&+\frac{ m_s [\langle \bar{s}s\rangle-2\langle \bar{q}q\rangle]}{96T^2}\langle\frac{\alpha_{s}GG}{\pi}\rangle\frac{1-x_i}{x_i}
-\frac{m_Q^2\langle \bar{s} s\rangle\langle \bar{q}q\rangle\pi^2}{108T^6}\langle\frac{\alpha_{s}GG}{\pi}\rangle\delta (s-m_Q^2)\nonumber\\
&&+\frac{\langle \bar{s} s\rangle\langle \bar{q}q\rangle\pi^2}{36T^4}\langle\frac{\alpha_{s}GG}{\pi}\rangle\delta (s-m_Q^2)\, ,
\end{eqnarray}
$x_i=\frac{m_Q^2}{s}$, the $T^2$ is the Borel parameter.

We derive the QCD sum rules in Eq.\eqref{QCDSR-I} in regard to $\frac{1}{T^2}$, then eliminate  the pole residues $\lambda_{+}$ and obtain the masses of the ground states $\Lambda_Q$ and $\Xi_Q$,
\begin{eqnarray}\label{QCDSR-I-der}
M_{+}^2&=&\frac{-\frac{d}{d(1/T^2)}\int_{m_Q^2}^{s_0}ds \Big[\sqrt{s}\rho_{QCD,1}(s)+\rho_{QCD,0}(s)\Big]\exp\left(-\frac{s}{T^2}\right)}{\int_{m_Q^2}^{s_0}ds \Big[\sqrt{s}\rho_{QCD,1}(s)+\rho_{QCD,0}(s)\Big]\exp\left(-\frac{s}{T^2}\right)}\, .
\end{eqnarray}
Thereafter, we will refer to the QCD sum rules in Eq.\eqref{QCDSR-I} and Eq.\eqref{QCDSR-I-der} as QCDSR I.

We   introduce the notations $\tau=\frac{1}{T^2}$, $D^n=\left( -\frac{d}{d\tau}\right)^n$, and  use the subscripts $1$ and $2$ to represent    the ground  states $\Lambda_Q$, $\Xi_Q$, and the first radially excited  states $\Lambda_Q(\rm 2S)$, $\Xi_Q(\rm 2S)$, respectively for simplicity.
\begin{eqnarray}\label{QCDSR-II-re}
\tilde{\lambda}_1^2\exp\left(-\tau M_1^2 \right)+\tilde{\lambda}_2^2\exp\left(-\tau M_2^2 \right)&=&\Pi^{\prime}_{QCD}(\tau) \, ,
\end{eqnarray}
where $\tilde{\lambda}_1^2=2M_{+}\lambda_{+}^2$, $\tilde{\lambda}_2^2=2M_{\rm 2S,+}\lambda_{\rm 2S,+}^2$,
 we introduce the subscript $QCD$ to denote  the QCD representation of the correlation functions  below the continuum thresholds $s_0^\prime$. Firstly, let us derive  the QCD sum rules in Eq.\eqref{QCDSR-II-re} with respect to $\tau$ to obtain,
\begin{eqnarray}\label{QCDSR-II-Dr}
\tilde{\lambda}_1^2M_1^2\exp\left(-\tau M_1^2 \right)+\tilde{\lambda}_2^2M_2^2\exp\left(-\tau M_2^2 \right)&=&D\Pi^{\prime}_{QCD}(\tau) \, .
\end{eqnarray}
From Eqs.\eqref{QCDSR-II-re}-\eqref{QCDSR-II-Dr}, we can obtain the QCD sum rules,
\begin{eqnarray}\label{QCDSR-II-Residue}
\tilde{\lambda}_i^2\exp\left(-\tau M_i^2 \right)&=&\frac{\left(D-M_j^2\right)\Pi^{\prime}_{QCD}(\tau)}{M_i^2-M_j^2} \, ,
\end{eqnarray}
where the sub-indexes $i \neq j$.
Then let us derive   the QCD sum rules in Eq.\eqref{QCDSR-II-Residue} with respect to $\tau$ to obtain
\begin{eqnarray}
M_i^2&=&\frac{\left(D^2-M_j^2D\right)\Pi_{QCD}^{\prime}(\tau)}{\left(D-M_j^2\right)\Pi_{QCD}^{\prime}(\tau)} \, , \nonumber\\
M_i^4&=&\frac{\left(D^3-M_j^2D^2\right)\Pi_{QCD}^{\prime}(\tau)}{\left(D-M_j^2\right)\Pi_{QCD}^{\prime}(\tau)}\, .
\end{eqnarray}
 The squared masses $M_i^2$ satisfy the  equation,
\begin{eqnarray}\label{Eq-abc}
M_i^4-b M_i^2+c&=&0\, ,
\end{eqnarray}
where
\begin{eqnarray}
b&=&\frac{D^3\otimes D^0-D^2\otimes D}{D^2\otimes D^0-D\otimes D}\, , \nonumber\\
c&=&\frac{D^3\otimes D-D^2\otimes D^2}{D^2\otimes D^0-D\otimes D}\, , \nonumber\\
D^j \otimes D^k&=&D^j\Pi^{\prime}_{QCD}(\tau) \,  D^k\Pi^{\prime}_{QCD}(\tau)\, ,
\end{eqnarray}
the indexes $i=1,2$ and $j,k=0,1,2,3$.
Finally we solve the equation in Eq.\eqref{Eq-abc} analytically to obtain two solutions \cite{Baxi-G,WangZG-4430},
\begin{eqnarray}\label{QCDSR-II-M1}
M_1^2&=&\frac{b-\sqrt{b^2-4c} }{2} \, ,
\end{eqnarray}
\begin{eqnarray}\label{QCDSR-II-M2}
M_2^2&=&\frac{b+\sqrt{b^2-4c} }{2} \, .
\end{eqnarray}
From  the QCD sum rules in Eqs.\eqref{QCDSR-II-M1}-\eqref{QCDSR-II-M2}, we can obtain both the masses of   the ground states and the first radial excited states, the   ground state masses from the QCD sum rules in Eq.\eqref{QCDSR-II-M1} suffer from additional uncertainties from the  first radial excited states $\Lambda_Q(\rm 2S)$ and $\Xi_Q(\rm 2S)$, and we neglect the QCD sum rules in Eq.\eqref{QCDSR-II-M1}. Thereafter, we will refer to the QCD sum rules in Eq.\eqref{QCDSR-II-Residue} and Eq.\eqref{QCDSR-II-M2} as the QCDSR II.

\section{Numerical results and discussions}
At the QCD side, we take the vacuum condensates  to be the standard values
$\langle\bar{q}q \rangle=-(0.24\pm 0.01\, \rm{GeV})^3$,  $\langle\bar{s}s \rangle=(0.8\pm0.1)\langle\bar{q}q \rangle$,
 $\langle\bar{q}g_s\sigma G q \rangle=m_0^2\langle \bar{q}q \rangle$, $\langle\bar{s}g_s\sigma G s \rangle=m_0^2\langle \bar{s}s \rangle$,
$m_0^2=(0.8 \pm 0.1)\,\rm{GeV}^2$, $\langle \frac{\alpha_s
GG}{\pi}\rangle=0.012\pm0.004\,\rm{GeV}^4$    at the energy scale  $\mu=1\, \rm{GeV}$
\cite{SVZ79,Reinders85,ColangeloReview}, and  take the $\overline{MS}$ masses $m_{c}(m_c)=(1.275\pm0.025)\,\rm{GeV}$, $m_{b}(m_b)=(4.18\pm0.03)\,\rm{GeV}$ and $m_s(\mu=2\,\rm{GeV})=(0.095\pm0.005)\,\rm{GeV}$
 from the Particle Data Group \cite{PDG}.
Moreover,  we take into account
the energy-scale dependence of  the quark condensates, mixed quark condensates and $\overline{MS}$ masses according to  the renormalization group equation,
 \begin{eqnarray}
 \langle\bar{q}q \rangle(\mu)&=&\langle\bar{q}q\rangle({\rm 1 GeV})\left[\frac{\alpha_{s}({\rm 1 GeV})}{\alpha_{s}(\mu)}\right]^{\frac{12}{33-2n_f}}\, , \nonumber\\
 \langle\bar{s}s \rangle(\mu)&=&\langle\bar{s}s \rangle({\rm 1 GeV})\left[\frac{\alpha_{s}({\rm 1 GeV})}{\alpha_{s}(\mu)}\right]^{\frac{12}{33-2n_f}}\, , \nonumber\\
 \langle\bar{q}g_s \sigma Gq \rangle(\mu)&=&\langle\bar{q}g_s \sigma Gq \rangle({\rm 1 GeV})\left[\frac{\alpha_{s}({\rm 1 GeV})}{\alpha_{s}(\mu)}\right]^{\frac{2}{33-2n_f}}\, ,\nonumber\\
  \langle\bar{s}g_s \sigma Gs \rangle(\mu)&=&\langle\bar{s}g_s \sigma Gs \rangle({\rm 1 GeV})\left[\frac{\alpha_{s}({\rm 1 GeV})}{\alpha_{s}(\mu)}\right]^{\frac{2}{33-2n_f}}\, ,\nonumber\\
m_b(\mu)&=&m_b(m_b)\left[\frac{\alpha_{s}(\mu)}{\alpha_{s}(m_b)}\right]^{\frac{12}{33-2n_f}} \, ,\nonumber\\
m_c(\mu)&=&m_c(m_c)\left[\frac{\alpha_{s}(\mu)}{\alpha_{s}(m_c)}\right]^{\frac{12}{33-2n_f}} \, ,\nonumber\\
m_s(\mu)&=&m_s({\rm 2GeV} )\left[\frac{\alpha_{s}(\mu)}{\alpha_{s}({\rm 2GeV})}\right]^{\frac{12}{33-2n_f}}\, ,\nonumber\\
\alpha_s(\mu)&=&\frac{1}{b_0t}\left[1-\frac{b_1}{b_0^2}\frac{\log t}{t} +\frac{b_1^2(\log^2{t}-\log{t}-1)+b_0b_2}{b_0^4t^2}\right]\, ,
\end{eqnarray}
  where $t=\log \frac{\mu^2}{\Lambda^2}$, $b_0=\frac{33-2n_f}{12\pi}$, $b_1=\frac{153-19n_f}{24\pi^2}$, $b_2=\frac{2857-\frac{5033}{9}n_f+\frac{325}{27}n_f^2}{128\pi^3}$,  $\Lambda=213\,\rm{MeV}$, $296\,\rm{MeV}$  and  $339\,\rm{MeV}$ for the flavors  $n_f=5$, $4$ and $3$, respectively  \cite{PDG,Narison-mix}.
For the charmed baryon states $\Lambda_c(\rm 1S,2S)$ and $\Xi_c(\rm 1S,2S)$, we choose the flavor numbers $n_f=4$, while for the bottom baryon states $\Lambda_b(\rm 1S,2S)$ and $\Xi_b(\rm 1S,2S)$, we choose the flavor numbers $n_f=5$.

In the QCDSR I, we choose the continuum threshold parameters to be  $\sqrt{s_0}=M_{gr}+0.50\pm0.10\,\rm{GeV}$ rather than to be $M_{gr}+0.6\sim0.8\,\rm{GeV}$ or $0.7\sim0.9\,\rm{GeV}$  as a constraint to exclude the contaminations from the first radial excited states \cite{M-Nielsen07,ZhangJR-2008,WangZG-EPJC-2010,WangZG-PLB}, where the subscript $gr$ denotes the ground states $\Lambda_Q$ and $\Xi_Q$.
Furthermore, we choose the energy scales of the QCD spectral densities in the QCD sum rules for the $\Lambda_c$, $\Xi_c$, $\Lambda_b$ and $\Xi_b$ to be the typical energy scales $\mu=1\,\rm{GeV}$, $1\,\rm{GeV}$, $2\,\rm{GeV}$ and $1.8\,\rm{GeV}$, respectively, where we subtract $0.2\,\rm{GeV}$ in the energy scale for the $\Xi_b$ to account for the finite mass of the $s$-quark.  After trial and error, we obtain the Borel parameters $T^2$, continuum threshold parameters $s_0$, pole contributions of the ground states and perturbative contributions, which are shown explicitly in Table \ref{Borel}.  From the Table, we can see that the pole contributions are about $(40-60)\%$ or $(40-70)\%$, the pole dominance is satisfied. The perturbative contributions are larger than $50\%$ except for the $\Lambda_b$, although  the perturbative contribution is about $(43-46)\%$ in that case, the contributions of the vacuum condensates of dimension 10 are tiny,  the operator product expansion is well convergent.

\begin{figure}
 \centering
 \includegraphics[totalheight=5cm,width=7cm]{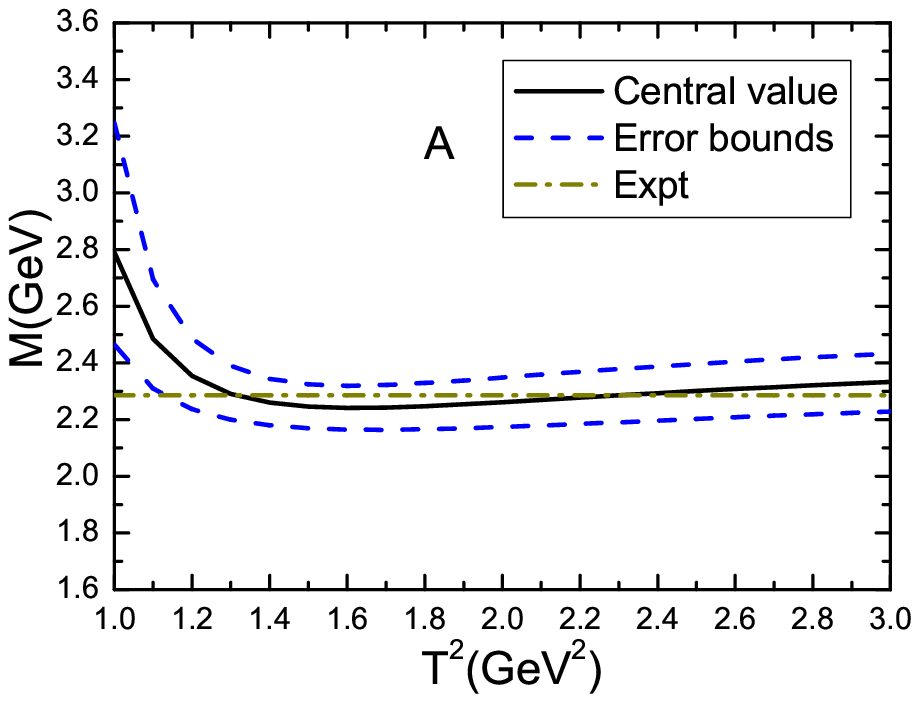}
 \includegraphics[totalheight=5cm,width=7cm]{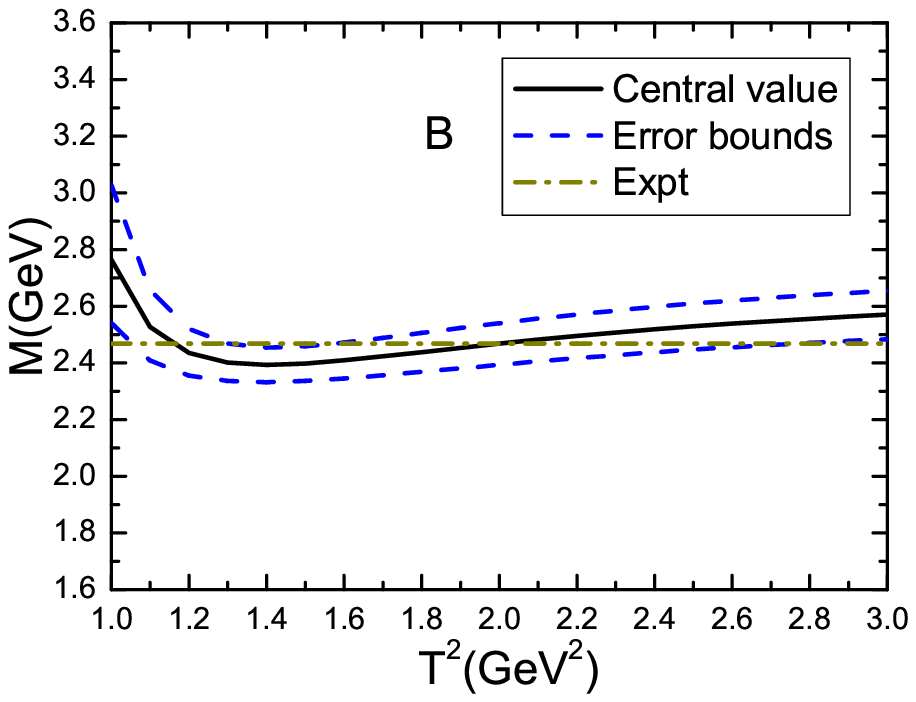}
 \includegraphics[totalheight=5cm,width=7cm]{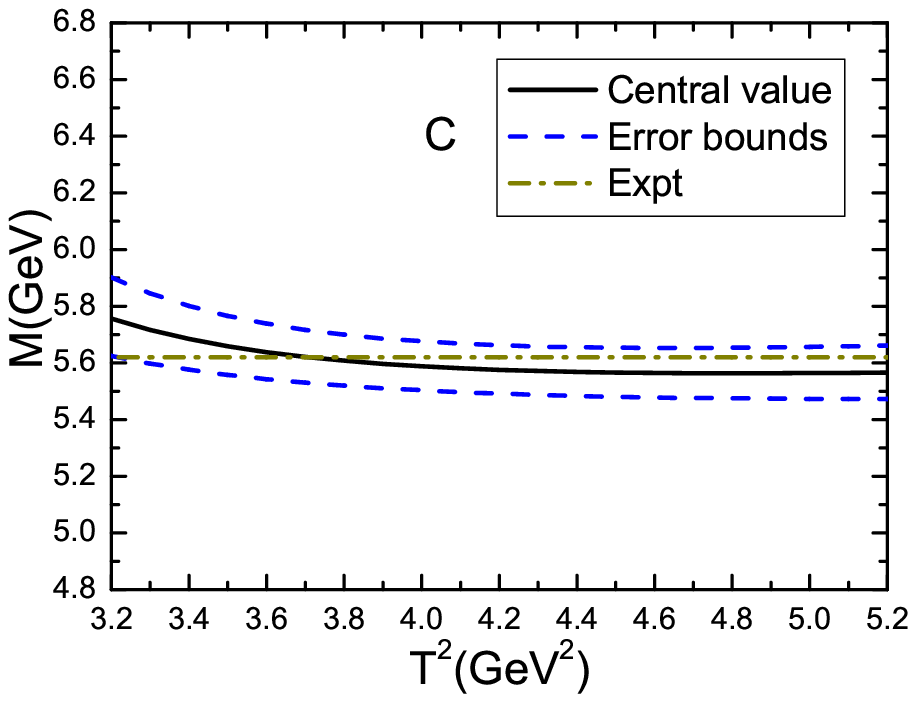}
 \includegraphics[totalheight=5cm,width=7cm]{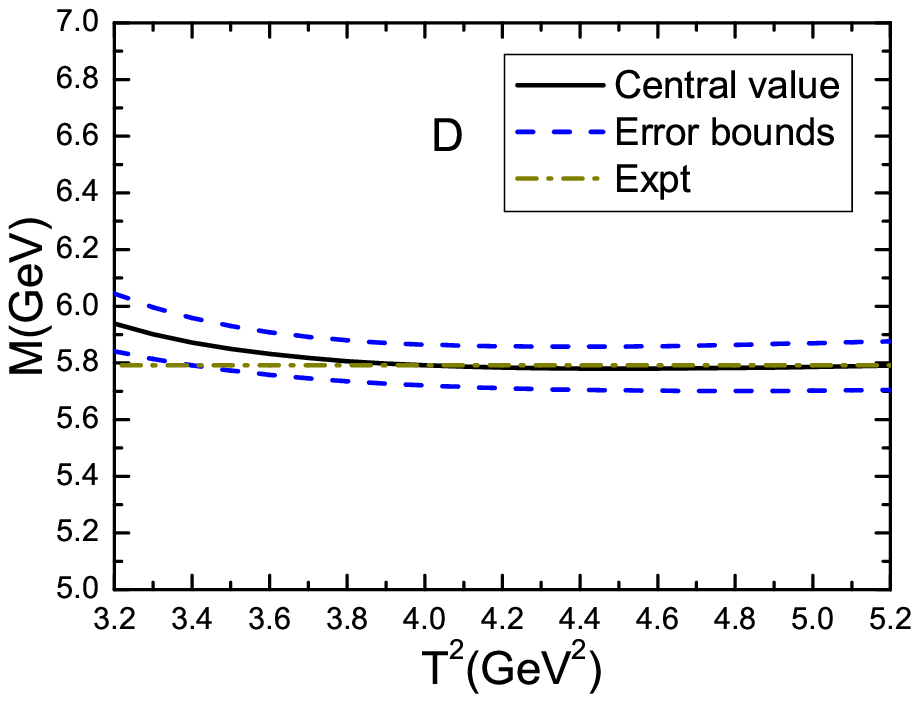}
  \includegraphics[totalheight=5cm,width=7cm]{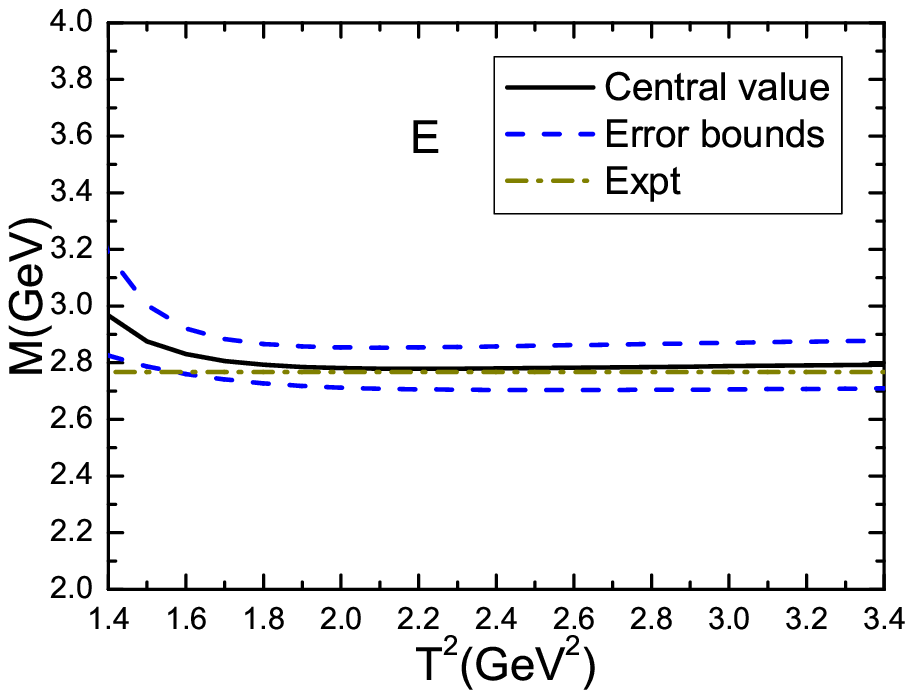}
 \includegraphics[totalheight=5cm,width=7cm]{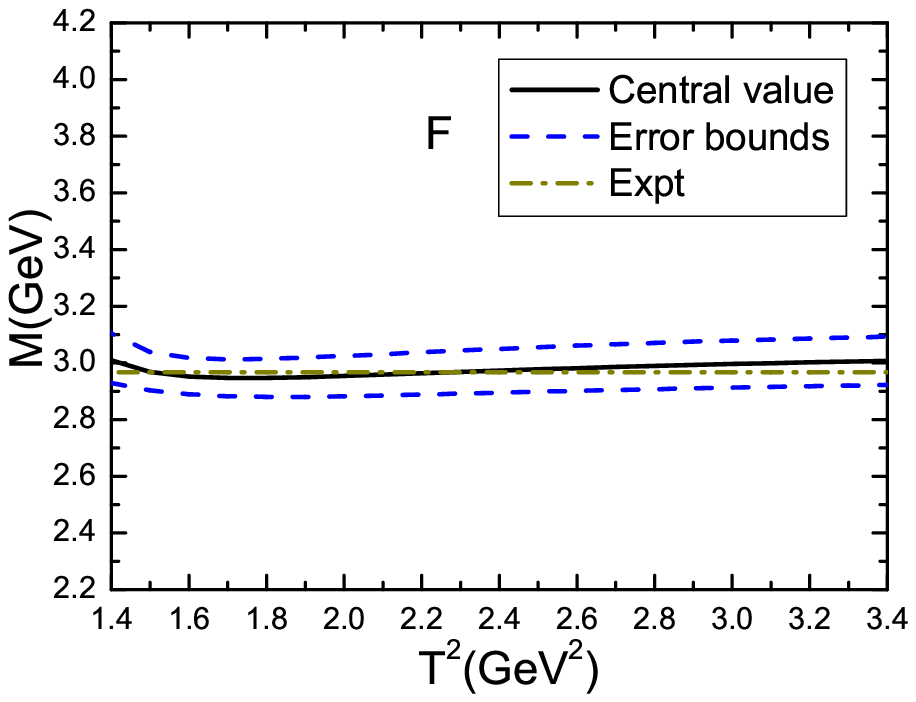}
 \includegraphics[totalheight=5cm,width=7cm]{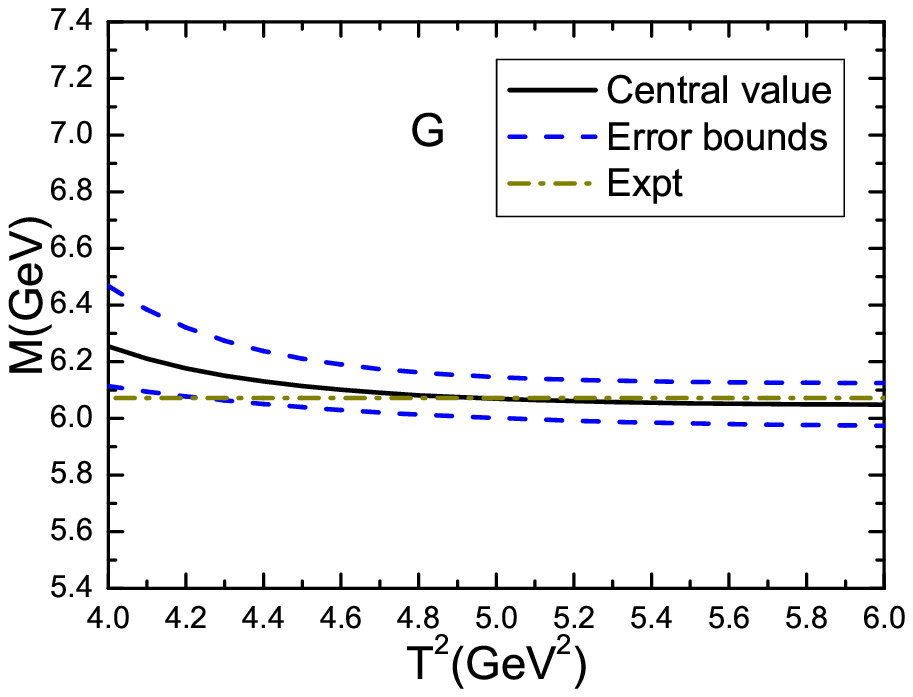}
 \includegraphics[totalheight=5cm,width=7cm]{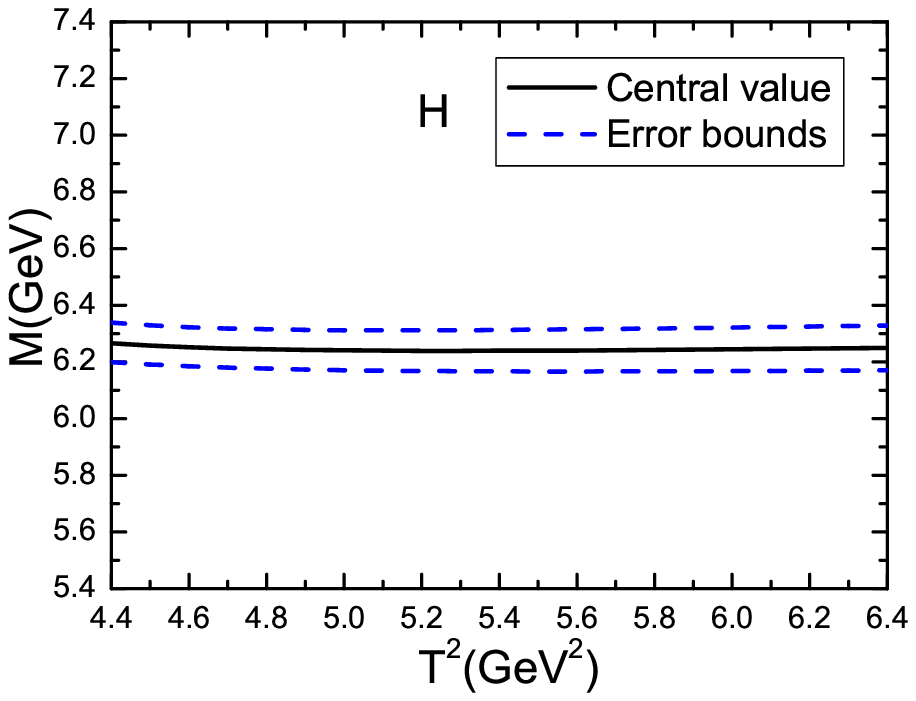}
 \caption{ The masses  with variations  of the Borel parameters $T^2$, where the $A$, $B$, $C$, $D$, $E$, $F$, $G$ and $H$ correspond to the $\Lambda_c$, $\Xi_c$, $\Lambda_b$, $\Xi_b$, $\Lambda_c(\rm 2S)$, $\Xi_c(\rm 2S)$, $\Lambda_b(\rm 2S)$ and $\Xi_b(\rm 2S)$, respectively, the expt denotes the experimental values.   }\label{mass-fig}
\end{figure}

\begin{figure}
 \centering
 \includegraphics[totalheight=5cm,width=7cm]{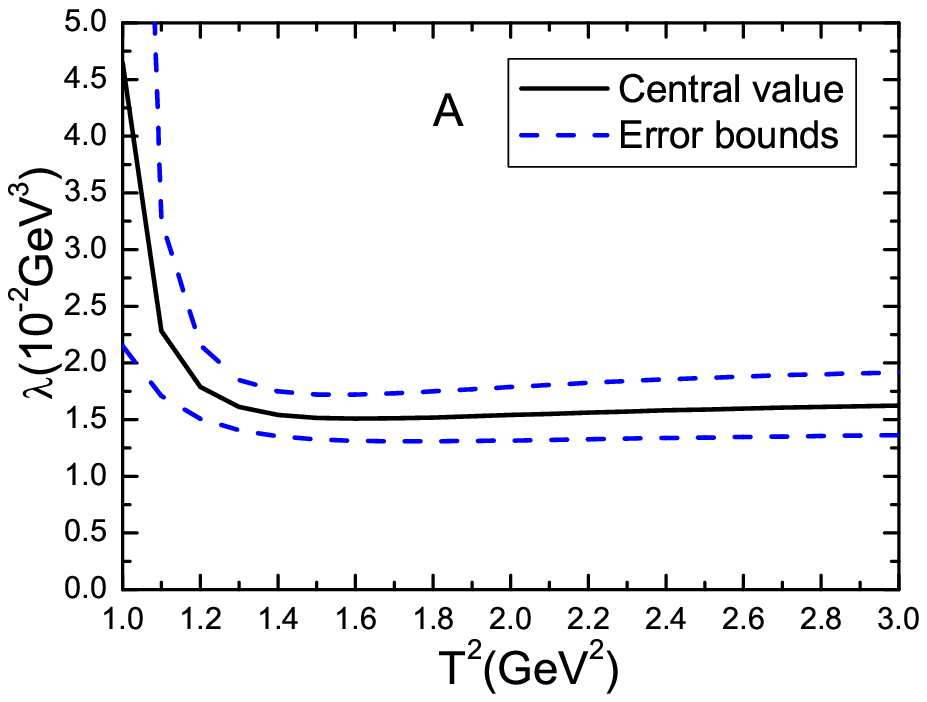}
 \includegraphics[totalheight=5cm,width=7cm]{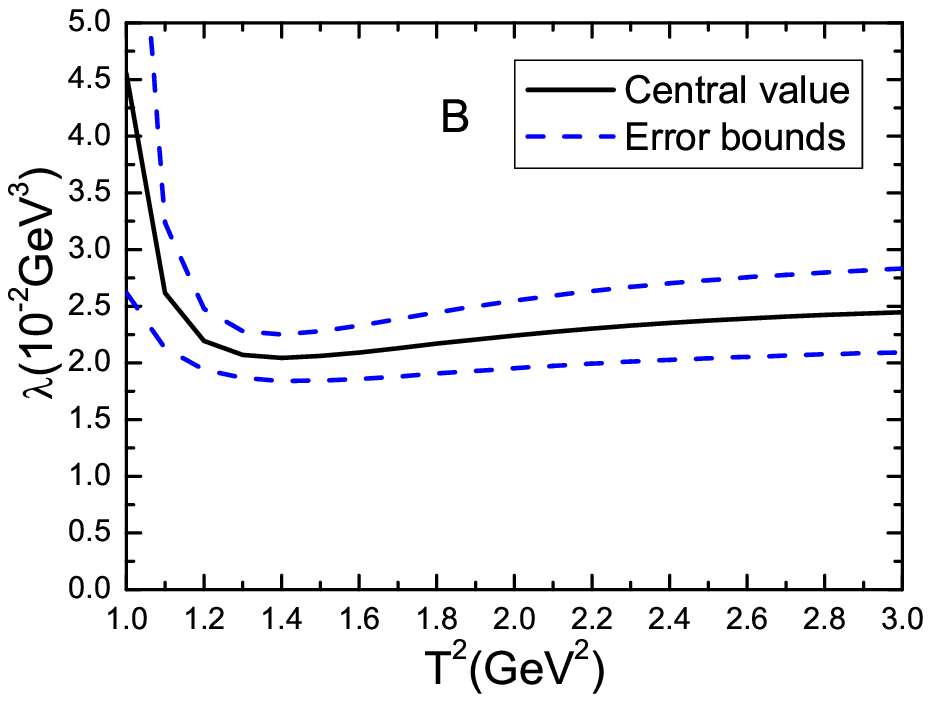}
   \includegraphics[totalheight=5cm,width=7cm]{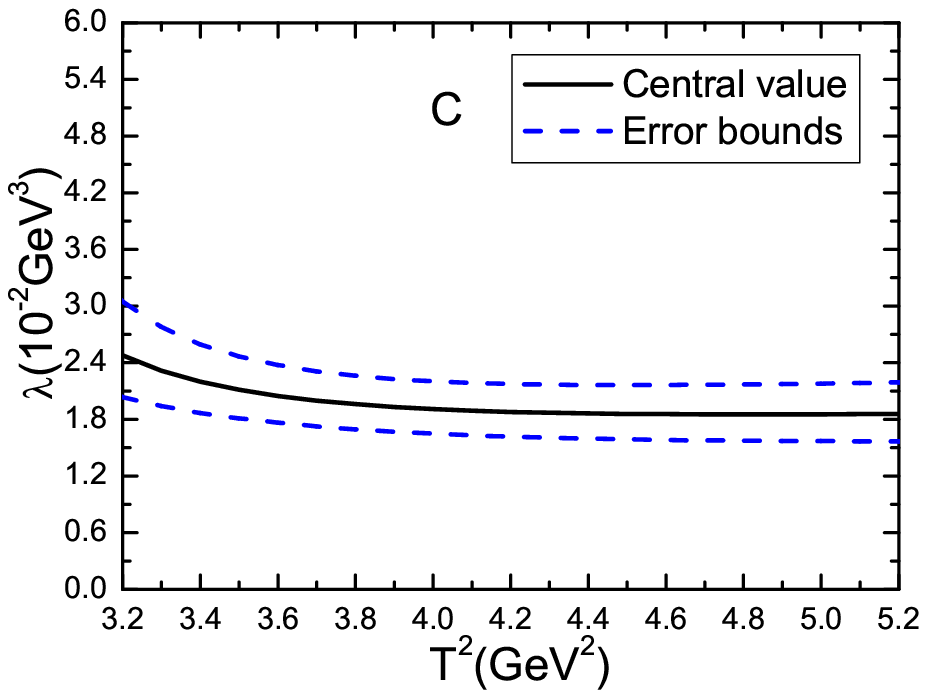}
 \includegraphics[totalheight=5cm,width=7cm]{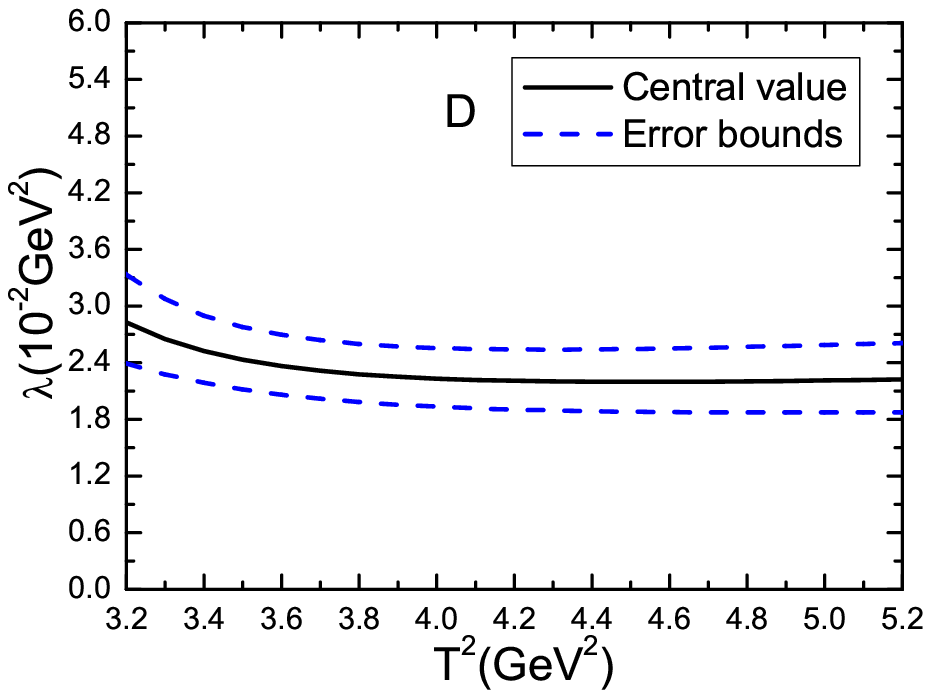}
 \includegraphics[totalheight=5cm,width=7cm]{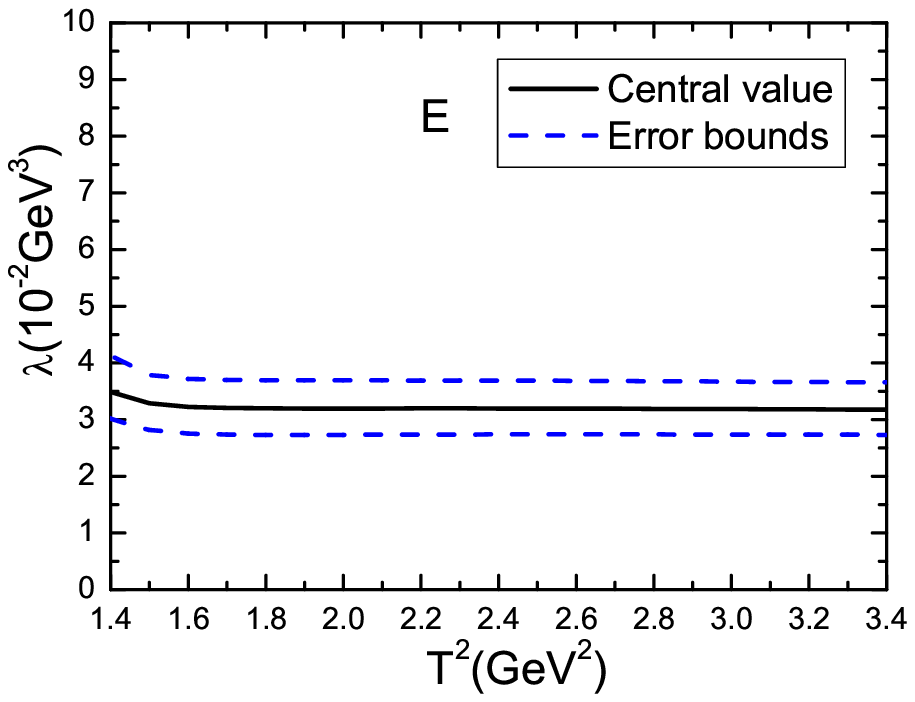}
 \includegraphics[totalheight=5cm,width=7cm]{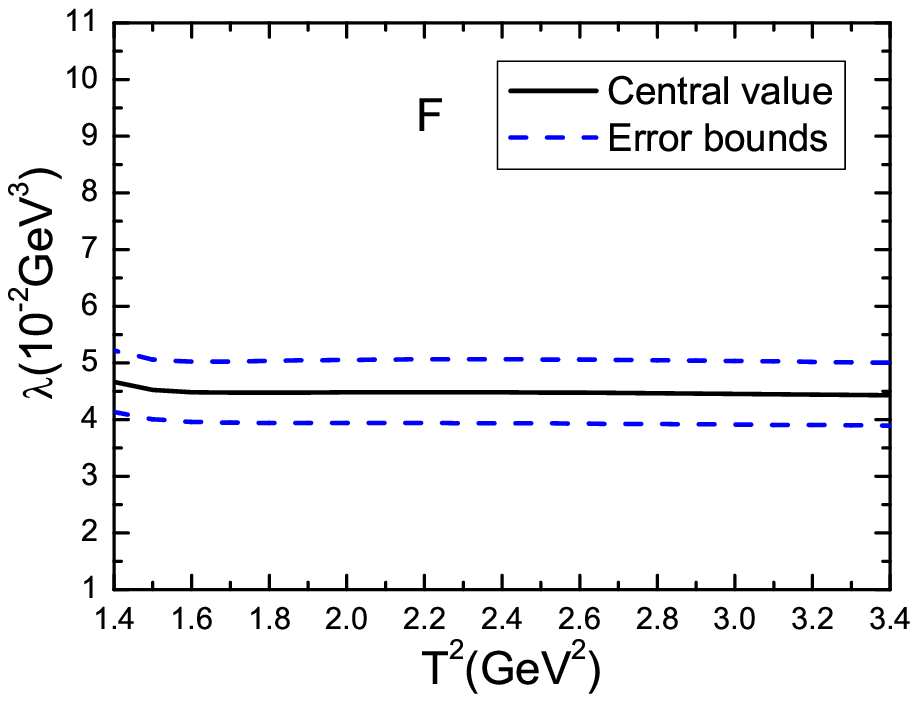}
  \includegraphics[totalheight=5cm,width=7cm]{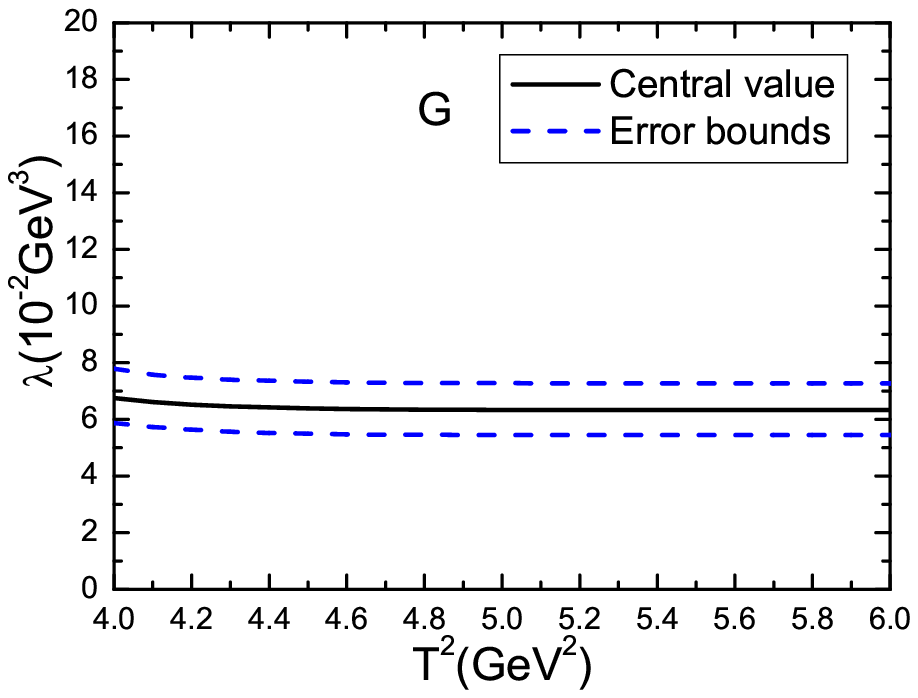}
 \includegraphics[totalheight=5cm,width=7cm]{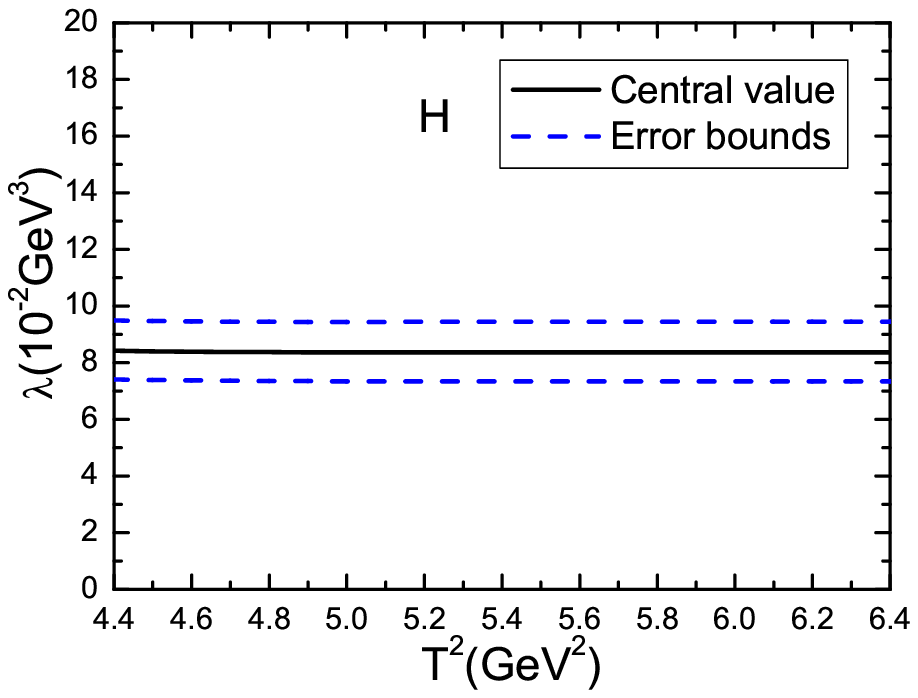}
 \caption{ The pole residues  with variations  of the Borel parameters $T^2$, where the $A$, $B$, $C$, $D$, $E$, $F$, $G$ and $H$ correspond to the $\Lambda_c$, $\Xi_c$, $\Lambda_b$, $\Xi_b$, $\Lambda_c(\rm 2S)$, $\Xi_c(\rm 2S)$, $\Lambda_b(\rm 2S)$ and $\Xi_b(\rm 2S)$, respectively.  }\label{residue-fig}
\end{figure}

\begin{figure}
 \centering
 \includegraphics[totalheight=7cm,width=10cm]{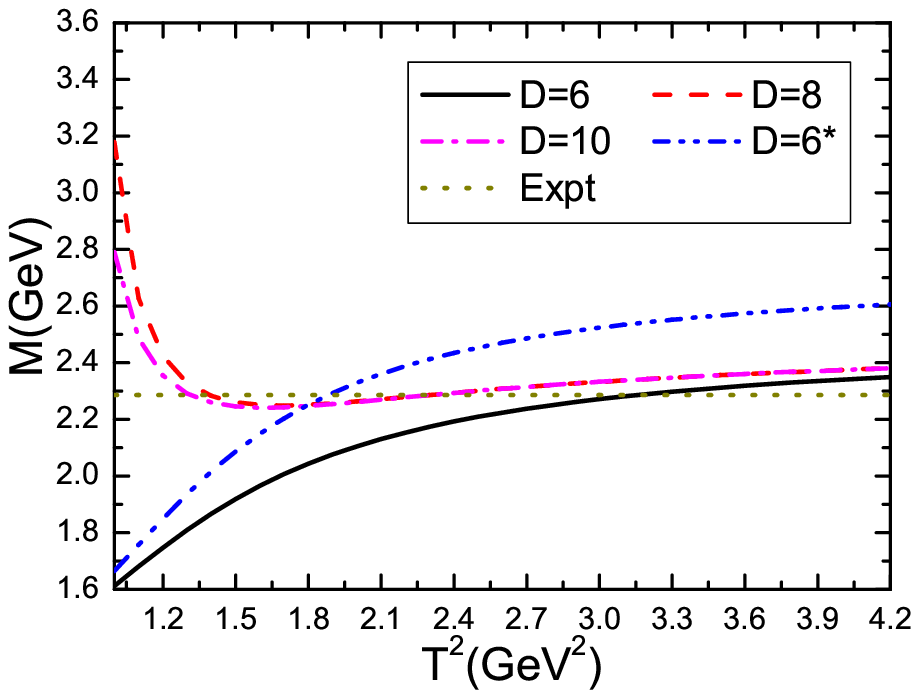}
  \caption{ The mass of the $\Lambda_c$  with variations  of the Borel parameter $T^2$, where the $D=6$, $8$ and $10$ denote  truncations of the vacuum condensates up to dimension $6$, $8$ and $10$, respectively, the star $*$ denotes the continuum threshold parameter  $\sqrt{s_0}=3.0\,\rm{GeV}$, the expt denotes the experimental value.  }\label{mass-D-fig}
\end{figure}

Now we take into account all uncertainties  of the input   parameters,
and obtain the values of the masses and pole residues of the ground states of the   flavor antitriplet  heavy baryon states $\Lambda_Q$ and $\Xi_Q$, which are
shown in Figs.\ref{mass-fig}-\ref{residue-fig} and Table \ref{mass}. From Table \ref{Borel} and Figs.\ref{mass-fig}-\ref{residue-fig}, we can see that there appear rather flat platforms in the Borel windows, the uncertainties originate from the Borel parameters are rather small. It is the first time that we obtain very flat platforms  for the heavy baryon states. From Tables \ref{Borel}-\ref{mass}, we can see that the central values have the relation $\sqrt{s_0}-M_{gr}=(0.50-0.55)\,\rm{GeV}$, the continuum threshold parameters $s_0$ are large enough to take into account all the ground state contributions   but small enough to suppress the first radial excited state contaminations sufficiently. Furthermore, they meet with our naive expectations.

In this article, we have neglected the perturbative  $\mathcal {O}(\alpha_s)$ corrections, if we  take into account the perturbative  $\mathcal {O}(\alpha_s)$ corrections, the perturbative terms  should be multiplied   by a factor
 $  1+C(s,m_Q^2)\frac{\alpha_s(T^2)}{\pi}  $,  where the $C(s,m_Q^2)$   are   some coefficients. Although we cannot estimate the uncertainties originate from the $\mathcal {O}(\alpha_s)$ corrections with confidence without explicit calculations, a crude estimation is still possible. In the case of the proton and neutron,  we can set $m_u=m_d=0$,
 and obtain the coefficient $C=\frac{53}{12}+\gamma_E$   \cite{Ioffe-2005}. If we take the approximation  $C(s,m_Q^2)=\frac{53}{12}+\gamma_E$, we can obtain the central values $M_{\Lambda_{c/b}}=2.29/5.64\,\rm{GeV}$ in stead of $2.24/5.61\,\rm{GeV}$, compared to the experimental values $2.29/5.62\,\rm{GeV}$ from the Particle Data Group \cite{PDG}, the central values $M_{\Lambda_{c/b}}=2.29/5.64\,\rm{GeV}$ are excellent. In fact, we should calculate the  perturbative  $\mathcal {O}(\alpha_s)$ corrections to the four-quark condensates $\langle\bar{q}q\rangle^2$ also, as they play an important role, and re-determine the Borel windows to extract the heavy baryon masses, just like in the case of the heavy mesons, where the perturbative  $\mathcal {O}(\alpha_s)$ corrections to the quark condensates $\langle\bar{q}q\rangle$ are also calculated  \cite{WangZG-Heavy-meson}. All in all, neglecting the perturbative  $\mathcal {O}(\alpha_s)$ corrections cannot impair the predictive ability remarkably, as we obtain the heavy baryon masses from fractions, the perturbative  $\mathcal {O}(\alpha_s)$ corrections in the numerators and denominators are canceled out with each other to a certain extent, see Eq.\eqref{QCDSR-I-der}.

In Fig.\ref{mass-D-fig}, we plot the predicted mass of the ground state $\Lambda_c$ with variations  of the Borel parameter $T^2$ by taking into account the vacuum condensates up to dimension 6, 8 and 10 respectively for the continuum threshold parameter $\sqrt{s_0}=2.75\,\rm{GeV}$.  From the figure, we can see that the truncation $D=6$ fails to
lead to a flat  platform and fails to reproduce the experimental value of the mass of the $\Lambda_c$, while the truncations $D=8$ and $10$ both lead to very flat platforms and reproduce the experimental value. In fact, the truncations $D=8$ and $10$ make tiny difference, which indicates that the vacuum condensates of dimension 8 (10) play an important (a tiny) role. We should take into account the vacuum condensates up to dimension 10 for consistence.  If we insist on taking the truncation $D=6$, we have to choose a much larger  continuum threshold parameter $\sqrt{s_0}=3.0\,\rm{GeV}$, the predicted mass increases  monotonically  with the increase of the Borel parameter $T^2$, we can reproduce  the experimental value of the mass of the $\Lambda_c$ with suitable Borel parameter but large uncertainty.

\begin{table}
\begin{center}
\begin{tabular}{|c|c|c|c|c|c|} \hline\hline
                        & $T^2 (\rm{GeV}^2)$   & $\sqrt{s_0}(\rm{GeV})$     & pole          & perturbative \\ \hline

 $\Lambda_c$            & $1.4-1.8$            & $2.75\pm0.10$              & $(40-72)\%$   & $(50-58)\%$\\ \hline

 $\Xi_c$                & $1.7-2.1$            & $3.00\pm0.10$              & $(42-71)\%$   & $(64-71)\%$\\ \hline

 $\Lambda_b$            & $3.6-4.0$            & $6.10\pm0.10$              & $(41-60)\%$   & $(43-46)\%$\\ \hline

     $\Xi_b$            & $3.8-4.2$            & $6.30\pm0.10$              & $(40-60)\%$   & $(51-54)\%$\\ \hline

 $\Lambda_c(\rm 2S)$    & $1.8-2.4$            & $3.00\pm0.10$              & $(41-74)\%$   & $(70-80)\%$\\ \hline

     $\Xi_c(\rm 2S)$    & $1.8-2.4$            & $3.25\pm0.10$              & $(54-84)\%$   & $(74-83)\%$\\ \hline

 $\Lambda_b(\rm 2S)$    & $4.6-5.0$            & $6.30\pm0.10$              & $(49-66)\%$   & $(76-79)\%$\\ \hline

     $\Xi_b(\rm 2S)$    & $5.1-5.5$            & $6.55\pm0.10$              & $(51-66)\%$   & $(83-85)\%$\\ \hline

\end{tabular}
\end{center}
\caption{ The Borel parameters $T^2$ and continuum threshold parameters $s_0(s_0^\prime)$
for the heavy baryon states, where the "pole" stands for the pole contributions from the ground states or the ground states plus the first radial excited states, and the "perturbative" stands for the contributions  from the perturbative terms.}\label{Borel}
\end{table}

\begin{table}
\begin{center}
\begin{tabular}{|c|c|c|c|c|c|c|}\hline\hline
                            & $M(\rm{GeV})$           & $\lambda(10^{-2}\rm{GeV}^3)$       & $M(\rm{GeV})[\rm{expt}]$  \\\hline

  $\Lambda_c$               & $2.24\pm0.09$           & $1.51\pm0.23$                      & 2.28646   \\ \hline

  $\Xi_c$                   & $2.45\pm0.10$           & $2.21\pm0.35$                      & 2.46795   \\ \hline

 $\Lambda_b$                & $5.61\pm0.12$           & $1.96\pm0.36$                      & 5.6196   \\ \hline

  $\Xi_b$                   & $5.79\pm0.09$           & $2.23\pm0.35$                      & 5.7919  \\ \hline

  $\Lambda_c(\rm 2S)$       & $2.78\pm0.08$           & $3.20\pm0.48$                      & 2.7666   \\ \hline

  $\Xi_c(\rm 2S)$           & $2.96\pm0.09$           & $4.48\pm0.56$                      & 2.9671    \\ \hline

$\Lambda_b(\rm 2S)$         & $6.08\pm0.09$           & $6.35\pm0.93$                      & 6.0723   \\ \hline

  $\Xi_b(\rm 2S)$           & $6.24\pm0.07$           & $8.36\pm1.05$                      &     \\ \hline

  $\Lambda_c(\rm 3S)$       &                         &                                    & 3.1749   \\ \hline

  $\Xi_c(\rm 3S)$           &                         &                                    & 3.3936    \\ \hline

$\Lambda_b(\rm 3S)$         &                         &                                    & 6.4935   \\ \hline

      \hline
\end{tabular}
\end{center}
\caption{ The masses  and pole residues  of the heavy baryon states, where the masses of the   $\Lambda_c(\rm 3S)$, $\Xi_c(\rm 3S)$ and
$\Lambda_b(\rm 3S)$ are obtained from the Regge trajectories.} \label{mass}
\end{table}

In the QCDSR II, we can borrow some ideas from the conventional charmonium states.
The  masses of the ground state, the first radial excited state and the second excited state of the charmonium states are $m_{J/\psi}=3.0969\,\rm{GeV}$,   $m_{\psi^\prime}=3.686097\,\rm{GeV}$ and $m_{\psi^{\prime\prime}}=4.039\,\rm{GeV}$ respectively from the Particle Data Group \cite{PDG}, the energy gaps are $m_{\psi^\prime}-m_{J/\psi}=0.59\,\rm{GeV}$, $m_{\psi^{\prime\prime}}-m_{J/\psi}=0.94\,\rm{GeV}$, we can  choose the continuum threshold parameters  $\sqrt{s_0^\prime}\leq M_{gr}+0.90\,\rm{GeV}$ tentatively  to avoid contaminations from the second radial excited states.
Furthermore, we choose the energy scales of the QCD spectral densities in the QCD sum rules for the $\Lambda_c(\rm 2S)$, $\Xi_c(\rm 2S)$, $\Lambda_b(\rm 2S)$ and $\Xi_b(\rm 2S)$ to be the typical energy scales $\mu=2\,\rm{GeV}$, $2\,\rm{GeV}$, $4\,\rm{GeV}$ and $3.8\,\rm{GeV}$, respectively, again we subtract $0.2\,\rm{GeV}$ in the energy scale for the $\Xi_b(\rm 2S)$ to account for the finite mass of the $s$-quark.  After trial and error, we obtain the Borel parameters $T^2$, continuum threshold parameters $s_0$, pole contributions and perturbative contributions, which are shown explicitly in Table \ref{Borel}.  From the Table, we can see that the pole contributions vary from  $40\%$ to $80\%$, the pole dominance is satisfied. The perturbative contributions are larger than $70\%$,  the operator product expansion is well convergent.

Again we take into account all uncertainties  of the input  parameters,
and obtain the values of the masses and pole residues of the first radial excited states of the   flavor antitriplet  heavy baryon states, which are also
shown in Figs.\ref{mass-fig}-\ref{residue-fig} and Table \ref{mass}. From Table \ref{Borel} and Figs.\ref{mass-fig}-\ref{residue-fig}, we can see that there appear rather flat platforms in the Borel windows, the uncertainties originate from the Borel parameters are rather small.
The predicted masses $M_{\Lambda_b(\rm 2S)}=6.08\pm0.09\,\rm{GeV}$, $M_{\Lambda_c(\rm 2S)}=2.78\pm0.08\,\rm{GeV}$ and $M_{\Xi_c(\rm 2S)}=2.96\pm0.09\,\rm{GeV}$,  are in excellent agreement with the experimental data $6072.3\pm 2.9\pm 0.6\pm 0.2\,\rm{MeV}$, $2766 .6\pm2.4\,\rm{MeV}$ and $2967 .1 \pm1.4\,\rm{MeV}$ \cite{LHCb-6072,PDG},  and support  assigning the $\Lambda_b(6072)$, $\Lambda_c(2765)$ and $\Xi_c(2980/2970)$ to be the first radial excited states of the $\Lambda_b$, $\Lambda_c$ and $\Xi_c$, respectively. The prediction $M_{\Xi_b(\rm 2S)}=6.24\pm0.07\,\rm{GeV}$ can be confronted to experimental data in the future.

If the masses of the ground states, the first radial excited states, the third radial excited states, etc of the heavy baryon states $\Lambda_Q$ and $\Xi_Q$ satisfy the  Regge trajectories,
 \begin{eqnarray}
 M_n^2&=&\alpha (n-1)+\alpha_0\, ,
 \end{eqnarray}
 with two parameters $\alpha$ and $\alpha_0$. We take the experimental values of the masses of the ground states and the first radial excited states shown in Table \ref{mass} as input parameters to fit the $\alpha$ and $\alpha_0$, and obtain the masses of the second radial excited states, which are also shown in Table \ref{mass} as the "experimental values".
 From the Tables \ref{Borel}-\ref{mass}, we can see that the continuum threshold parameters $\sqrt{s_0^\prime}\leq M_{\Lambda_c(\rm 3S)}$, $M_{\Xi_c(\rm 3S)}$ and $M_{\Lambda_b(\rm 3S)}$, respectively,  the contaminations from the second radial excited states are excluded.
 The central values have the relations $\sqrt{s^{\prime}_0}-M_{\rm 2S}=(0.20-0.30)\,\rm{GeV}$ and $M_{\rm 3S}-\sqrt{s^{\prime}_0}=(0.15-0.20)\,\rm{GeV}$, the continuum threshold parameters $s^\prime_0$ are large enough to take into account all the first radial excited state contributions   but small enough to exclude the second radial excited state contaminations. The central values $\sqrt{s^{\prime}_0}-M_{gr}=(0.70-0.80)\,\rm{GeV}$, which are consistent with the experimental value $m_{\psi^{\prime\prime}}-m_{J/\psi}=0.94\,\rm{GeV}$ \cite{PDG}.

In Ref.\cite{LuQF-6072}, the Liang and Lu study the strong decay behaviors under various
assignments of the $\Lambda_b(6072)$ within the ${}^3P_0$ model, and obtain the conclusion that the $\Lambda_b(6072)$ can be assigned to be
  the $\rho$-mode excitation of the $\Lambda_b$ family with the spin-parity $J^P={\frac{1}{2}}^-$ by introducing the mixing effects between the $s_l=0$ and $s_l=1$ states, where the $s_l$ denotes the angular momentum of the light degrees of freedom. Accordingly, we can introduce the relative P-wave between the $u$ and $d$ quarks explicitly and construct the current $J(x)$ to interpolate the $\Lambda_b(6072)$,
\begin{eqnarray}
J(x)&=&J_0(x) \cos\theta+J_1(x) \sin\theta\, \nonumber\\
J_0(x)&=&\varepsilon^{ijk}  u^T_i(x)C\gamma^\alpha \stackrel{\leftrightarrow}{\partial}_\alpha d_j(x)   b_k(x)\, ,\nonumber\\
J_1(x)&=&\varepsilon^{ijk}  u^T_i(x)C\gamma_\alpha \stackrel{\leftrightarrow}{\partial}_\beta d_j(x) \sigma^{\alpha\beta}  b_k(x)\, ,
\end{eqnarray}
where $\stackrel{\leftrightarrow}{\partial}_\alpha=\overrightarrow{\partial}_\alpha-\overleftarrow{\partial}_\alpha$. Without direct calculating the mass and decay width, we cannot obtain the conclusion whether or not the QCD sum rules support such an assignment, this is our next work.

The spin-parity of the ground states $\Lambda_c$, $\Xi_c$, $\Lambda_b$ and $\Xi_b$ have been established, the values listed in the Review of Particle Physics are $J^P={\frac{1}{2}}^+$ \cite{PDG}. In this article, we study the masses and pole residues of the ground states and the first radial excited states of the flavor
antitriplet heavy baryons, and make possible assignments of the   $\Lambda_b(6072)$, $\Lambda_c(2765)$ and $\Xi_c(2980/2970)$ according to the predicted masses, as their spin-parity have not been established yet.
The present predictions support assigning the $\Lambda_b(6072)$, $\Lambda_c(2765)$ and $\Xi_c(2980/2970)$ to be the first radial excitations     of the $\Lambda_b$, $\Lambda_c$ and $\Xi_c$, respectively, more theoretical and experimental works are still needed to make more reliable assignments. There is no experimental candidate for the $\Xi_c(\rm 2S)$ state. After the manuscript was submitted to
https://arxiv.org, and appeared as arXiv:1704.01854, the Belle collaboration determined the spin-parity of the $\Xi_c(2970)^+$ to be ${\frac{1}{2}}^+$ for the first time \cite{Belle-Xi-2S-JP}, which is consistent with the present calculation.

\section{Conclusion}
In this article, we construct the $\Lambda$-type currents to study  the ground states and the first radial excited states of the  flavor antitriplet heavy baryon states $\Lambda_Q$ and $\Xi_Q$ with the spin-parity  $J^P={1\over 2}^{+}$ by subtracting the contributions
from the corresponding  heavy baryon states with the spin-parity $J^P={1\over 2}^{-}$ via
the QCD sum rules. We carry out the operator product expansion up to the vacuum condensates of dimension $10$ in a consistent  way, and observe that the higher  dimensional vacuum condensates play an important role, and obtain very stable QCD sum rules with variations of the Borel parameters for the ground states for the first time.  Then  we study the masses and pole residues of the first radial excited states in details,  the predicted masses $M_{\Lambda_b(\rm 2S)}=6.08\pm0.09\,\rm{GeV}$, $M_{\Lambda_c(\rm 2S)}=2.78\pm0.08\,\rm{GeV}$ and $M_{\Xi_c(\rm 2S)}=2.96\pm0.09\,\rm{GeV}$  are in excellent agreement with the experimental data,  and support  assigning the $\Lambda_b(6072)$, $\Lambda_c(2765)$ and $\Xi_c(2980/2970)$ to be the first radial excited states of the $\Lambda_b$, $\Lambda_c$ and $\Xi_c$, respectively.  Finally we use the   Regge trajectories to obtain the masses of the second radial excited states and observe that the continuum threshold parameters are reasonable to avoid the contaminations from the second radial excited states.

\section*{Acknowledgements}
This  work is supported by National Natural Science Foundation, Grant Number  11775079.

\end{document}